\documentstyle[epsfig]{article} 

\title{
BACKGROUND DISCRIMINATION CAPABILITIES OF A HEAT AND 
IONIZATION GERMANIUM CRYOGENIC DETECTOR\\
\vskip1truecm
\small
{\it EDELWEISS COLLABORATION}
}

\author{
P. Di Stefano$^{\rm a}$\thanks{Present address: Max-Planck-Institut f\"ur Physik,
 F\"ohringer Ring 6, D-80805 Munich, Germany},
L. Berg\'e$^{\rm b}$, B. Chambon$^{\rm c}$, 
M. Chapellier$^{\rm d}$, J. Chaumont$^{\rm b}$,\\
G. Chardin$^{\rm a}$, P. Charvin$^{\rm a, e}$, M. De J\'esus$^{\rm c}$, D. Drain$^{\rm c}$, L. Dumoulin$^{\rm b}$,\\
P. Forget$^{\rm d}$,
P. Garoche$^{\rm f}$, 
J. Gascon$^{\rm c}$, C. Goldbach$^{\rm g}$
D. L'H\^ote$^{\rm d}$,\\ J. Mallet$^{\rm a}$,
J. Mangin$^{\rm h}$, S. Marnieros$^{\rm b}$,
L. Miramonti $^{\rm a}$,
L. Mosca $^{\rm a}$,\\ X-F. Navick $^{\rm a}$,
G. Nollez $^{\rm g}$, P. Pari $^{\rm d}$, S. P\'ecourt$^{\rm c}$,
E. Simon$^{\rm c}$,\\
L. Stab $^{\rm i}$,
J-P. Torre $^{\rm j}$, R. Tourbot $^{\rm d}$,
D. Yvon $^{\rm a}$\\
\\
\small
{\it$^{\rm a}$ CEA, Centre d'Etudes Nucl\'eaires de Saclay, DSM/DAPNIA, F-91191 Gif-sur-Yvette Cedex, France}\\
\small
{\it$^{\rm b}$ CSNSM, IN2P3-CNRS, Universit\'e Paris XI, b\^at. 108, F-91405 Orsay Cedex, France}\\
\small
{\it$^{\rm c}$ IPN Lyon and UCBL, IN2P3-CNRS, 43 Bd. du 11 novembre 1918, F-69622 Villeurbanne Cedex, France}\\
\small
{\it$^{\rm d}$ CEA, Centre d'Etudes Nucl\'eaires de Saclay, DSM/DRECAM, F-91191 Gif-sur-Yvette Cedex, France}\\
\small
{\it$^{\rm e}$ Laboratoire Souterrain de Modane, CEA-CNRS, 90 rue Polset, F-73500 Modane, France}\\
\small
{\it$^{\rm f}$ Laboratoire de Physique des Solides, Universit\'e Paris XI, F-91405 Orsay Cedex, France}\\
\small
{\it$^{\rm g}$ Institut d'Astrophysique de Paris, INSU-CNRS, 98 bis Bd. Arago, F-75014 Paris, France}\\
\small
{\it$^{\rm h}$ LPUB, Universit\'e de Bourgogne, F-21078 Dijon, France}\\
\small
{\it$^{\rm i}$ Institut de Physique Nucl\'eaire, Universit\'e Paris XI, F-91405 Orsay Cedex, France}\\
\small
{\it$^{\rm j}$ Service d'A\'eronomie, BP 3, F-91371 Verri\`eres le Buisson Cedex, France}\\
\\
\normalsize
}

\date{\vspace{-0.7cm}}

\begin{document}

\maketitle

\begin{abstract}
The discrimination capabilities of a 70 g heat and ionization Ge bolometer are studied. This first prototype
has been used 
by the EDELWEISS Dark Matter experiment, installed in the Laboratoire Souterrain de Modane, for direct detection of WIMPs. 
Gamma and neutron calibrations demonstrate that this type of detector is able to reject more than 99.6\% of the 
background while retaining 95\% of the signal, provided that the background events distribution is not biased towards the 
surface of the Ge crystal. However, the 1.17 kg.day of data taken in a relatively important radioactive environment show an 
extra population slightly overlapping the signal. This background is likely due to interactions of low energy photons or 
electrons near the surface of the crystal, and is somewhat reduced  by applying a higher  charge-collecting inverse bias voltage  
( -6 V instead of -2 V) to the Ge diode. Despite this contamination, more than 98\% of the background
 can be rejected while retaining 50\% of the signal.
This yields a conservative upper limit of 0.7 ${\rm event.day^{-1}.kg^{-1}.keV_{recoil}^{-1}}$ at 90\% confidence level 
in the 15-45 keV recoil energy interval; the present sensitivity appears to be
limited by the fast ambient neutrons.
Upgrades in progress on the installation are summarized.

\end{abstract}

\newpage

\section{Introduction}
 	The search for dark matter is one of the major challenges of contemporary physics. Despite 
extensive scans for MAssive Compact Halo Objects (MACHOs) initiated by several groups \cite{alc,aub,pac}, it appears at 
present that the local density of dark matter can hardly be composed of baryonic matter essentially. Indeed, recent 
results from these experiments indicate that a significant fraction of the MACHOs observed in the direction of the 
Small Magellanic Cloud (SMC) are probably due to deflectors internal to the SMC, further reducing the possible 
amount of dark matter in the form of MACHOs in the halo itself \cite{alc2}.

The coincidence of the electroweak interaction scale (SUSY theory) with that required for 
Weakly Interacting Massive Particles (WIMPs) to contribute significantly to the solution of the dark matter problem is a further 
motivation for attacking the WIMPs direct detection challenge. In addition these hypothetical particles 
represent an attractive solution to the problem of galaxy formation. Experimental efforts are underway to detect 
these particles, either indirectly, by searching for products of their annihilation in the core of the Sun or of the 
Earth, or directly, by detecting the interactions of the WIMPs themselves in ordinary matter, as first suggested in 
the mid 1980s \cite{stodol,good} .

The sensitivity of present experiments appears to be limited by the radioactive background rate of the 
detectors \cite{baud1}  and by the systematics of the rejection scheme using statistical identification methods based on pulse 
shape discrimination techniques (PSD) in NaI crystals \cite{dama,ukdmc,gerb} . The motivation for the development of cryogenic 
detectors is the possibility, when the heat measurement is
coupled with a measurement of ionization \cite{shutt,schne,drain,berge} (or of 
scintillation \cite{meunier}), to discriminate much more reliably the main source of radioactive background, producing 
electron recoils, from the nuclear recoils expected from WIMP interactions. Indeed, WIMPs 
scatter off nuclei, which ionize (or produce scintillation) less efficiently 
than electron recoils for a given energy deposit 
in the absorber. 

    	 In the following, we present the discrimination capabilities of this new generation of detectors in the 
EDELWEISS (Exp\'erience pour DEtecter Les Wimps En SIte Souterrain)  underground experiment and we show 
that a 70 g high-purity Ge crystal can already reach sensitivity levels typical of the best existing experiments despite its
small mass and a relatively high radioactivity level in the detector environment. The present limitations of these 
detectors are then explored and the first attempts to counteract them are
discussed.

\section{The EDELWEISS experiment}
The EDELWEISS cryogenic experiment operates in the Laboratoire Souterrain de Modane (LSM), an 
underground laboratory off the Fr\'ejus highway tunnel beneath the French-Italian Alps. The 4600 mwe of rock 
reduce the muon flux to one part in two million of its surface value, that is to about 4.5 muons.m$^{-2}$.day$^{-1}$. The fast ambient neutron flux is measured to be $4.\ 10^{-6} {\rm cm^{-2}.s^{-1}}$ \cite{veren}. 

The cryogenic setup and  shielding are presented in \cite{adb}. 
In particular, the dilution refrigerator has a cooling power of
100 $\mu$W at 100 mK, a useful volume for bolometer installation of about one
liter and can reach temperatures as low as 10 mK. The optimized readout electronics are described in \cite{yvon}.
The results presented here 
concern a prototype germanium detector of only moderate radioactive cleanliness, without near Roman lead 
shielding, radon removal or ambient neutron moderator.

\subsection{The cryogenic detector}

	The absorber of the detector consists in a 70 g germanium monocrystal of high purity ($n_D-n_A\cong 5.10^9 {\rm cm}^{-3}$). 
The crystal shape is a disc 8 mm thick and 48 mm in diameter with bevelled edges.

	Both charge and heat signals resulting from a particle interaction are measured by this detector. Charge 
collection is achieved by an electric field applied between two electrodes realized by boron implantation on one face 
of the monocrystal and by phosphorus implantation on the other. The applied bias voltages are of the order of a few 
volts. The {\it p-i-n} structure has been chosen to decrease as much as possible the injection of carriers in the detector volume and to allow large
bias voltages.
The doses of implanted ions are two orders of magnitude above the insulator-metal transition. The implantation
energies have been chosen in order to reduce the thickness of the non-metallic implanted region 
and to ensure that the metallic layer reaches the detector surface. Recrystallization in the implanted region has been 
achieved by a fast thermal annealing and its quality checked by ellipsometry and by SIMS.\footnote{Second Ionisation Mass Spectroscopy. These electrode
characterizations have been done with a previous detector by J.-P. Ponpon of
the PHASE Laboratory (Strasbourg, France).}
In addition, sheet 
resistances have been measured on this last detector to verify the metallic behavior at very low temperatures 
\cite{lhot1,xfn}. Several improvements of the detector design result from a study of the trap ionization
and neutralization mechanisms in the bulk of the detector. The very low density of ionized traps
we reached allows a noticeable improvement of the ionization channel time stability and energy resolution [20-24].
 In order to minimize edge effects and to maximize charge collection, the thickness of the monocrystal 
has been reduced to 4 mm along its contour to increase the field strength, thereby reducing trapping and 
recombination.

	The heat measurement is based on the variation, according to the 
Mott-Efr\"os-Schklovskii law, of the resistivity of a Neutron Transmutation Doped Germanium (NTD). This thermal sensor
  is glued to one face of the detector. The size of this NTD is 2x1x0.8 mm$^3$. The heat sink consists of three copper wires 50 $\mu$m in diameter
and about 2 cm in length. The operating temperature of the NTD 
sensor is about 20 mK for a mixing chamber temperature around 10 mK.

 The whole detector lies on sapphire balls to ensure a good thermal decoupling from the detector 
holder made of low radioactivity copper and brass. A detailed description of the detector manufacturing and performances is given in [19-24].

\subsection{Off-line data analysis}

 The small size of the heat and ionization signals (typically 
 a temperature increase of $\approx 10^{-6}$ K, and of the order of a thousand electron-hole pairs) exposes 
them to various electronic noises, be they fundamental or instrumental (Figure 1). Therefore, analysis of the data 
has been performed off-line using various methods including the optimal filtering technique in Fourier space \cite{mose} . 
This involves a least-square fit in frequency space between the actual event and a model event (obtained by 
averaging away the noise on several real events) while inversely weighting each frequency by the standard 
deviation of the noise at that frequency. 

Neglecting pileup and assuming pulse shapes independent of pulse amplitude, each event $s$ can be represented as the sum
of a scaled and shifted model $m$ and of a random noise $n$, $s(t)=Am(t-t_s)+n(t)$. Further assuming the noise to
be gaussian in frequency space, we can construct a $\chi^2$ as the weighted 
difference between signal and
scaled and shifted model \cite{mose}:
$$ \chi^2=\sum_{\omega}{|N(\omega)|^2\over \sigma_{\omega}^2}=\sum_{\omega}{|S(\omega)-Ae^{-i\omega t_s}M(\omega)|^2
\over \sigma_{\omega}^2}, $$
where $\sigma_{\omega}$ is the standard deviation of the noise at frequency $\omega/2\pi$ and capital letters
represent the Fourier transform of their lowercase counterparts.
Minimization with respect to $A$ and $t_s$ 
yields an estimate of these two parameters.  
Parseval's identity allows us to simply maximize over time shifts the scalar product $A=\sum_t f(t-t_s)s(t)$
where the function $f$ can be calculated from its Fourier transform once for
all events \cite{szym}
$$F(\omega)={{e^{-i\omega t_s}M(\omega)\over \sigma_{\omega}^2}}\bigg / {\sum_\omega {|M(\omega)|^2\over\sigma_{\omega}^2}}.$$ 
Calculating the $\chi^2$ to judge the quality of the fit is however more tractable in frequency space, and this 
remains computationally feasible thanks to Fast Fourier Transforms. 

With respect to time fit techniques, this method selectively filters out noisy frequencies and thus is 
advantageous in the many cases of non-white noise spectra (Figure 1). Like straightforward time fit techniques 
however, it is ill-suited to data presenting pileups - though given current masses of bolometers this is not yet a problem
when looking for WIMPs.

\section{In situ detector calibrations}

\subsection{ Energy normalization and recoil energy determination}

The heat signal results from the sum of the heat deposited by the incident particle and of the heat 
generated by the charge carriers during their drift (the so-called Luke-Neganov effect \cite{luke}). This last term is given 
by the product of the bias voltage times the collected charge. Thus as bias voltage increases, the fraction of the heat signal directly correlated to 
the ionization signal increases, hampering separation of the electron recoils
from the nuclear recoils. In practice, although bias voltages up 
to -12 V have been used, most of the data have been accumulated under bias voltages of -2 V and -6 V. There 
is in fact
an additional electric field to that applied; it is caused by the 0.7 eV gap in Ge which creates a -0.7 V 
reverse field in the {\it p-i-n} diode. However, only the applied field ultimately counts for the Luke-Neganov effect : the 
energy equal to the additional field times the collected charge is indeed released when the charges are collected, but 
merely compensates the energy used to create the pairs in the first place \cite{schutt2}.

Moreover it is standard procedure to normalize heat 
(phonon) energies (keV heat) to equivalent electron energies (keV e.e.). 
Volume electron recoils (with almost complete charge collection) are used to calibrate the charge and phonon channels.
For such a volume electron recoil of a given energy (e.g. 122 keV for a $^{57}$Co calibration), both charge
and phonon amplitudes ($E_{ch}$ and $E_{ph}$ respectively)
are normalized to the deposited energy. In order to reconstruct this energy deposited
in the detector, or recoil energy $E_{rec}$, from these quantities, the following formula \cite{luke,dis}
is applied:
$$E_{rec}=E_{ph}(1+{eV\over\epsilon_\gamma})-E_{ch}{eV\over\epsilon_\gamma},$$
where $V$ is the bias voltage for charge collection (in Volts), $e$ the elementary charge and $\epsilon_\gamma$ the average energy per 
electron-hole pair (in eV) for an electron recoil.
The recoil energy resolution
becomes thus bias-dependent, as shown in Figure 6.
In the absence of Luke-Neganov effect
($V=0$) $E_{rec}=E_{ph}$; for a 
photon interaction where all the charge is collected (volume electron recoil) $E_{rec}=E_{ph}=E_{ch}$.

	It is shown in \cite{dis,ben} how the $\epsilon_\gamma$  parameter can be measured using, in X- and $\gamma$-ray
line calibrations, the 
small fraction of events which exhibits an incomplete charge collection. The measured value is close to the standard 
value  $\epsilon_\gamma\cong$ 3 eV for the 77 K germanium diode detectors \cite{knoll}.

\subsection{Electron recoil thresholds and resolutions}

Electron recoil energy resolutions and thresholds have been studied using a $^{57}$Co source which provides
essentially photoelectric interactions from its main peak at 122 keV. The ionization signal being roughly 1000 
times faster than the heat signal, the former has usually been used 
 as trigger for the acquisition with an effective threshold 
at 4 keV equivalent electron (e.e.) on the charge channel. Threshold efficiency has been studied by comparing a 
$^{60}$Co calibration with a Monte Carlo simulation. It is found to rise rapidly to 100\% by 6 keV \cite{lino}, see Figure 2.

The heat and ionization channels exhibit energy FWHM resolutions of $\approx$ 1
and $\approx$ 1.2 keV e.e. at 122 keV. According to the E$_{rec}$ expression 
of section 3.1, the corresponding resolutions on the recoil energy are
$\approx$ 1.8 keV at -2 V and $\approx$ 3.8 keV at -6 V.

\subsection{Signal and background calibrations}

The detector has been calibrated with both neutron and gamma sources, in order to check its power of 
separation between nuclear recoil ("signal") and electron recoil (background) events. High energy photons 
provide Compton interactions throughout the crystal rather than biased towards its surface.

   	 A  $^{252}$Cf neutron source, of about 2.5 $\mu$Ci activity, has been used for this calibration, at two different values 
of the reverse bias voltage (-2 V and  -6 V). The branching ratio for neutron emission in the decay of the 
$^{252}$Cf isotope is about 12\%  and the average kinetic energy of the emitted neutrons is around 2 MeV.  The $^{252}$Cf 
isotope is also a gamma source with a branching ratio of about 60\%, and about 80\% of these gammas have an 
energy less than 1 MeV. So, in principle this source alone is sufficient  to obtain simultaneously both neutron and 
gamma calibrations. In order to improve the statistics without  too long an exposure of the detector to the neutron 
source and the associated risk of neutron activation, we use also a $^{60}$Co source to mimick the background noise of electron recoils. Results are 
presented in Figure 3 for both polarization voltages, showing a remarkably good separation between gamma and neutron
recoil events down to low energy. While most electron recoil events appear well behaved, and neatly line 
up along the main diagonal in the heat-ionization plane, approximately 5\% suffer from incomplete charge 
collection.

	The recoil energy $E_{rec}$ is derived according to section 3.1. 
Merging the data obtained at the two 
bias voltages yields the quenching factor $(E_{ch}/E_{rec})$ versus $E_{rec}$ in Figure 4.
The curve $4{\rm keV}/E_{rec}$  , which  represents the threshold (at 4 keV 
e.e.), is also given. A "neutron line" (the 
quenching factor for nuclear recoils) can be parameterized using the measured spreads and mean values of the 
charge over recoil ratio, calculated in discrete recoil energy intervals above 20 keV (under 20 keV a threshold bias 
is clearly visible). We obtain $E_{ch}/E_{rec} = (0.16\pm0.07)E_{rec}^{0.18\pm0.1}$, a result in agreement (within error limits) with 
those of the Heidelberg-Moscow    collaboration \cite{baud2}, parameterized in 
\cite{baud1} by
$E_{ch} = 0.14 E_{rec}^{1.19}$, and in turn in 
agreement with the Lindhard theory \cite{lind} and most of the previous experiments (see \cite{baud2}
and references therein).
The comparison with the results of the CDMS experiment \cite{shutt}, displayed in the Figure 4 inset, shows a very good agreement.

       A neutron zone of known acceptance can be constructed by adjusting a power 
law of the recoil energy to the points at +$n\sigma$,-$m\sigma$ above and 
below the mean charge over recoil ratio. For a given background, the choice
of the $n$ and $m$ values is a matter of optimizing the signal-to-noise ratio.
On the Figure 4, the 95\% acceptance neutron zone ($\pm 2 \sigma$) is
displayed as an example. This neutron calibration
induces a slight bias in the determination of the quenching factor: the non-negligible number of multiple diffusions in the crystal
 leads to a merging of the quenching factor values at different energies.
The width of the nuclear recoil band is overestimated accordingly for WIMPs
for which it should be defined by single interactions only.

We now turn our attention to the separation capabilities of the detector based upon the two types of 
calibrations just discussed.  
In Figure 5 we apply the neutron zone method to the data of the $^{60}$Co
calibration taken at -2 V. In the 15-100 keV recoil energy interval, 3 events 
(out of 3229) are found within the 95\%-acceptance nuclear recoil band: (99.90
$\pm$ 0.05) \% of the photon background is rejected. At -6V we find a rejection of (99.6$\pm$ 0.2) \%. These results are very promising; however it must be 
kept in mind that they have been obtained from calibrations and that 
operational values may differ as we will see later.

\section{A realistic case of data taking}

Following the benchmarks just discussed, 0.65 kg.day and 1.17 kg.day of data were taken at the bias 
voltages of -2 V and -6 V respectively. Both spectra show a rate of roughly
$\approx 35-40 \ {\rm event.kg^{-1}.day^{-1}.keV_{recoil}^{-1}}$ in the 
15-45 keV recoil energy interval, below the 46.5 keV peak due to $^{210}$Pb contamination.The recoil energy versus ionization
plots of the data (Figure 6) show a large number of off-axis events.
This population, interpreted as incomplete charge collection of electron surface 
events, is discussed in \cite{ben} together with another population which appears at recoil energies of the 
order of 80-100 keV and  is attributed to  surface nuclear recoils.

Given the large number of events with incomplete charge collection, 
we only retain the (0$\sigma$, -2$\sigma$) neutron zones as discussed in
the previous section (Figure 7). This improves the signal-to-noise ratio and
halves the acceptance to 0.475. The samples recorded respectively at -2 V and -6 V bias 
contain 6 events (respectively 7 events) in the 15-45 keV recoil range. We 
assume they are nuclear recoils, and  use them to derive a conservative upper limit on a signal. For the respective 
exposures and acceptance, they  yield upper limits at the 90\% confidence level of 1.1 ${\rm event.kg^{-1}.day^{-1}.keV_{recoil}^{-1}}$  at  
-2 V, and 0.7 ${\rm event.kg^{-1}.day^{-1}.keV_{recoil}^{-1}}$ at -6 V.
It will appear below that a large fraction of the
events observed in the nuclear recoil band
 can be attributed to neutron interactions inside the 
detector. One can thus only set a lower limit of 98\% for the rejection factor of the electron recoil background.

When we compare the mean ionization over recoil energy ratio of the off-axis events (electron surface 
events) at both bias voltages ($0.46\pm0.01$  at -2 V and  $0.539\pm0.006$  at -6 V), we find that the higher bias
voltage brings the off-axis events significantly closer to the well-collected events. 

We have also run the detector using a heat trigger. Because the ionization 
signal is roughly 1000 times faster than the heat signal, it is not always
possible to correlate the heat pulse with the charge pulse. Nonetheless,
this trigger mode has brought to light a distinct second type of thermal
event, one with a rise time dominated by electronics and practically never correlated
with an ionization event. For these two reasons we attribute these events to
interactions in the heat sensor itself. The energy spectrum of these events is compatible with a small amount of tritium activity in the NTD thermometer.

The events surviving in the neutron zone are not sufficiently separated from the 
population of electron surface events to be attributed with certainty to nuclear recoil events, induced in the Ge 
detector by ambient neutrons (the observed rate being  much higher than what would be expected from 
WIMPs\cite {baud1,dama}). The flux and energy spectrum of the fast ambient neutrons originating for the most part from internal 
radioactivity in the surrounding rock, have been measured in the LSM by our collaboration using conventional 
scintillation detectors. The flux was found to be  $4.0\pm0.1\  10^{-6}{\rm cm}^{-2}.{\rm s}^{-1}$ (statistical error only) \cite{veren}.
A Monte-Carlo simulation was then performed 
to predict the expected number of neutron-induced nuclear recoils in our 70 g Ge detector.
 It uses the GEANT Monte-Carlo simulation \cite{brun} in conjunction 
with a package specially developed \cite{kerr} to handle low energy 
neutrons (from a few eV to a few MeV).
The expected and measured numbers of recoils in the nuclear recoil band (47.5 \% acceptance) over the recoil 
energy interval 15 - 100 keV are given in Table 1.
\begin{table}[h]
\begin{tabular}{|c|c|c|c|}
\hline
Bias voltage & Exposure (kg.day) & Measured recoil number & Simulated neutron number\\ 
\hline\hline
-2 V & 0.65 & $14\pm4$ & $7\pm2$\\
\hline
-6 V & 1.17 & $15\pm4$ & $12\pm3$\\
\hline
\end{tabular} 
\caption{Measured number of events and simulated number of ambient neutron events in
 the nuclear recoil band (47.5 \% acceptance), over 15-100 keV recoil energy. The error bars on the simulated neutron numbers
result from the experimental error on the measured neutron flux.}
\end{table}

By inspection of  Table 1 it appears that at the -6 V polarization a large fraction, and possibly all the 
events observed in the nuclear recoil region can be attributed to neutron interactions inside the 
detector. At  -2 V however, the separation between nuclear recoils and electron surface events is, as already noted, 
slightly less favorable, and accordingly the simulation  indicates a pollution of the nuclear recoil band by 
surface events.

\section{Conclusion and outlook}

	Our prototype heat and ionization bolometer provides an excellent separation of the calibrated 
background and signal down to an energy of 4 keV e.e.. 
However data taken during background runs contain an important population of off-axis events (electron surface events) essentially not seen during the calibrations. Despite this, the rejection capabilities of the 
detector remain satisfactory, still reducing the number of electron recoil events by nearly two orders of magnitude. Sensitivity limits on 
WIMPs \cite{drain,berge} obtained with this prototype are only a factor of 10 higher than the best existing limits obtained 
with longer-established techniques \cite{baud1,dama} despite the small mass (70 g) of our detector and the relatively high 
radioactivity level in the detector environment. Moreover the present sensitivity of the experiment appears to be 
limited, principally for the runs with a -6V bias voltage, not by the gamma background or by parasitic electron surface events, but by the ambient neutrons  in the 
underground laboratory. Given the systematic uncertainties in the Monte Carlo neutron simulations  a background 
subtraction is not attempted but, clearly, shielding against neutrons is required.

	A new generation of detectors is being tested in the LSM, with  new implantation  schemes for the 
electrodes and a rigorous selection procedure of the crystal holder material 
\cite{XFN3}.
Several improvements have 
been undertaken in the detector environment: nitrogen flushing for radon removal, improved passive protection against 
gamma radioactivity (including  close Roman lead shielding) and 30 cm removable paraffin shielding against neutrons. 
In an initial test run, two 70 g prototypes show background levels before rejection of  
$\approx 2\  {\rm event.keV^{-1}.kg^{-1}.day^{-1}}$  below 50 keV, a 
factor of 10 improvement over previous data, with a comparable progression for electron surface events. The number of 
events remaining in the neutron zone is now extremely limited and could be due to residual neutron 
interactions inside the present shielding. Assuming a good performance of the 
neutron shielding and that the energy resolution and background rejection are 
the 
same as those reported here, a few months of data taking with these detectors should allow a test of the entire 
parameter region claimed by the DAMA experiment \cite{dama} and the entrance 
into the relevant supersymmetry parameter space.

\section{Acknowledgments}
The support of the technical staff of the Laboratoire Souterrain de Modane
 and of the participating laboratories is 
gratefully acknowledged.
We thank J.P. Passerieux, A. Le Coguie and B. Cahan for the quality of their
contribution to the drawing and realization of electronic components.
 This work has been partially funded by the EEC-Network program under contract 
ERBFMRXCT980167.

\bibliography{}

\begin{thebibliography}{99}  

\bibitem{alc}
C. Alcock et al. (MACHO Collaboration) Nature 365 (1993) 621.

\bibitem{aub}
E. Aubourg et al. (EROS Collaboration) Nature 365 (1993) 623.

\bibitem{pac}
B. Paczynski et al., Bull. Am. Astron. Soc.  187 (1995) 1407.

\bibitem{alc2}
C. Alcock et al. (EROS and MACHO Collaborations) , Ap. J. 499 (1998) L9.

\bibitem{stodol}
A. Drukier and L. Stodolsky, Phys. Rev. D 30 (1984) 2295.

\bibitem{good}
M. W. Goodman and E. Witten, Phys. Rev. D 31 (1985) 3059.

\bibitem{baud1}
L. Baudis et al., Phys. Rev. D 59 (1999) 022001.

\bibitem{dama}
R. Bernabei et al., Phys. Lett. B389 (1996) 757 ;
Phys. Lett. B424 (1998) 195 ; Phys. Lett. B 450 (1999) 448.

\bibitem{ukdmc}
P.F.Smith et al., Phys. Lett. B379 (1996) 299 ; Phys. Rep. 307 (1998) 275.

\bibitem{gerb}
G. Gerbier et al., Astrop. Phys. 11 (1999) 287. 

\bibitem{shutt}
T. Shutt et al., Phys. Rev. Lett. 69 (1992) 3425.

\bibitem{schne}
R.W. Schnee et al., Phys. Rep. 307 (1998) 283.

\bibitem{drain}
D. Drain et al., Phys. Rep. 307 (1998) 297.

\bibitem{berge}
L. Berg\'e et al., Nucl. Phys. B 70 (1999) 69.

\bibitem{meunier}
P. Meunier et al., Appl. Phys. Lett. 75 (1999) 1335.

\bibitem{veren}
V. Chazal et al., Astrop. Phys. 9 (1998) 163.

\bibitem{adb}
A. de Bellefon et al., Astrop. Phys. 6 (1996) 35.

\bibitem{yvon}
D. Yvon et al., Nucl. Instr. Methods Phys. Res. A 368 (1996) 778.

\bibitem{lhot1}
D. L'H\^ote et al., Nucl. Instr. Methods Phys. Res. A 370 (1996) 193.

\bibitem{xfn}
X.-F. Navick, Th\`ese de Doctorat, No 97/PA07/7146, Universit\'e Paris VII (1997), unpublished .

\bibitem{lhot2}
X.-F. Navick et al., Proc. of the 2nd Conf. on new developments in 
photo-detection, Beaune, 1999, in press in Nucl. Instr. Methods Phys.
Res A.

\bibitem{xfn2}
X.-F. Navick et al., Nucl. Instr. Methods Phys. Res. A370 (1996) 213;
Proc. 7th Int. Workshop on Low Temperature Detectors, Munich, Germany, Ed. S. Cooper (1997) 244.

\bibitem{lhot3}
D. L'H\^ote et al., Czechoslovak  J. of  Phys.  46-S5 (1996) 2903.

\bibitem{lhot4}
D. L'H\^ote et al., J. Appl. Phys. 87 (2000) 1507.

\bibitem{mose}
S.H. Moseley, J.C. Mather and D. Mc Cammon, J. Appl. Phys. 56 (1984) 1257.

\bibitem{szym}
A.E. Szymkowiak, R.L. Kelley, S.H. Moseley and C.K. Stahle, J. Low Temp. Phys. 93 (1993) 281.

\bibitem{luke}
P.N. Luke, J. Appl. Phys. 64 (1988) 6858.

\bibitem{schutt2}
T. Shutt et al., Phys. Rev. Lett. 69 (1992) 3531.

\bibitem{dis}
P. Di Stefano, Th\`ese de Doctorat, Universit\'e Paris XI Orsay (1998), unpublished.

\bibitem{ben}
A. Benoit et al., to be published in Phys. Letters B.

\bibitem{knoll}
G. E. Knoll, Radiation Detection and Measurement, (J. Wiley, New York, 1989).

\bibitem{lino}
L. Miramonti, Th\`ese de Doctorat, Universit\'e Paris XI Orsay (1999), unpublished.

\bibitem{baud2}
L. Baudis et al., Nucl. Instr. Methods Phys. Res. A 418 (1998) 348.

\bibitem{lind}
J. Lindhard, V. Nielsen, M. Scharff and P.V. Thomsen, Mat. Fyz. Medd. Dan. 
Vid. Selsk. \\ 33 (1963) 1.

\bibitem{brun}
GEANT Detector Description and Simulation Tool, CERN Program Library,W5103,
CERN (1993).

\bibitem{kerr}
H. de Kerret and B. Lefi\`evre, Report, Coll\`ege de France, LPC 88 01, (1988), unpublished.

\bibitem {XFN3}
X.-F. Navick et al., Proc. of th 8th Intern. Workshop on Low Temperature 
Detectors, to appear in Nucl. Instr. Methods Phys. Res. A.  

\end{thebibliography}

\begin{figure}[ht]
\epsfig{file=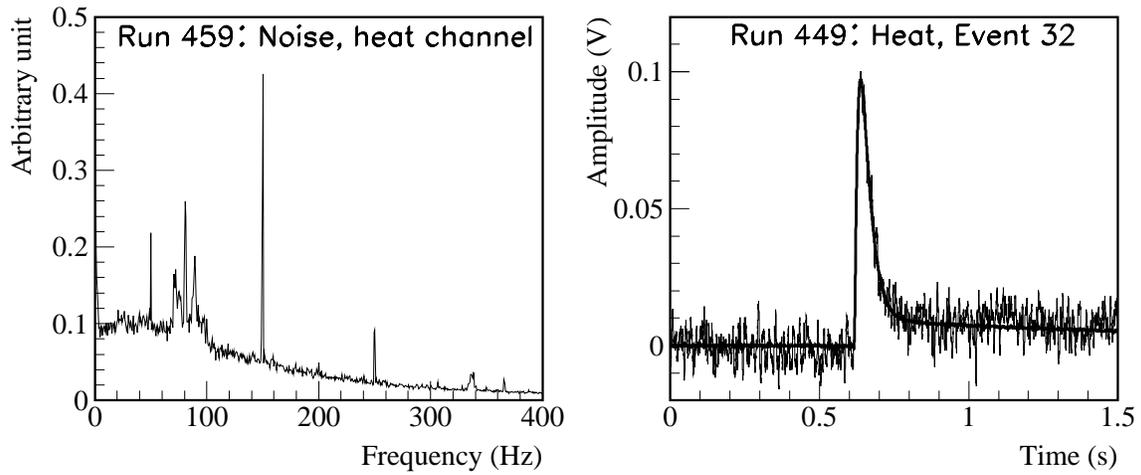,width=17cm}
\caption[]{Left: a noise spectrum of the heat channel. Overall shape
is that of thermodynamical noise level of the NTD, squashed by low-pass
filter. Two types of peaks are visible: narrow electromagnetic ones due
essentially to the odd harmonics of the electrical supply, and wider ones
due to microphonics. Right: result of optimal
filter fit on a heat event ($\approx$ 60 keV).}
\label{fig1}
\end{figure}
\begin{figure}[hb]
\epsfig{file=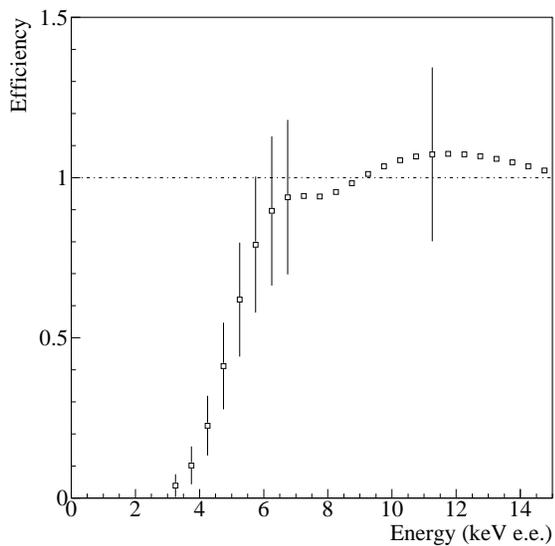,width=8cm}
\caption[] {The threshold efficiency of the ionization channel, determined
by comparing the data of $^{60}$Co calibrations with a Monte Carlo simulation.}
\label{fig2}
\end{figure}
\begin{figure}[ht]
\epsfig{file=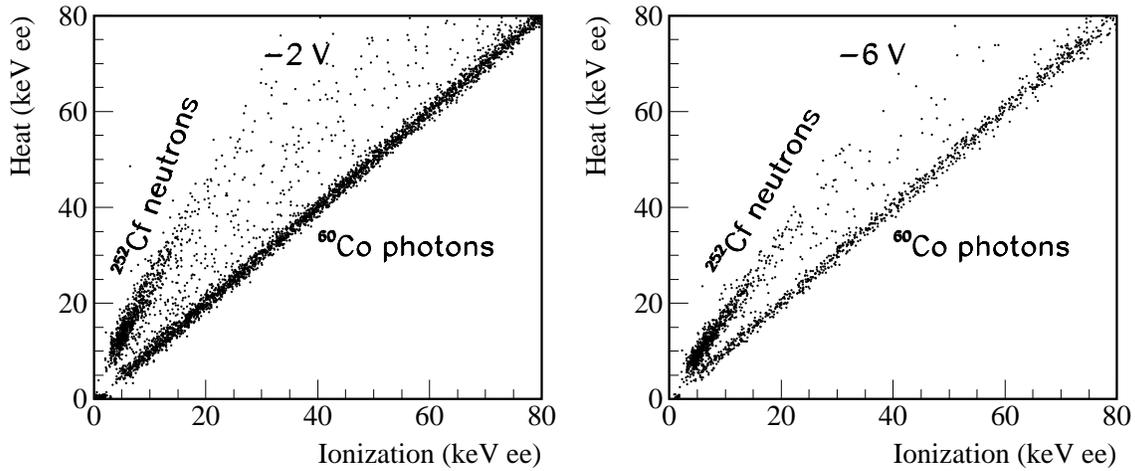,width=17cm}
\caption[] {Heat-ionization planes of $^{60}$Co and $^{252}$Cf calibrations
at two bias voltages. Over the 0 - 80 keV e.e. energy interval, 5\% of the events from $^{60}$Co are 
ionization-deficient at -2 V and 3\% at -6 V.}
\label{fig3}
\end{figure}
\begin{figure}[hb]
\epsfig{file=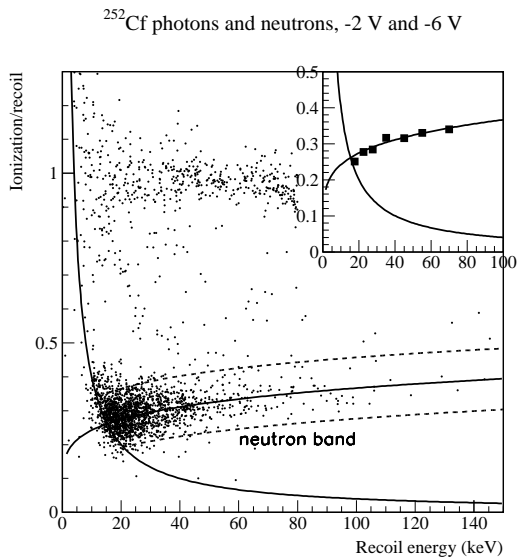,width=8cm}
\caption[] {The quenching factor, or ionization over recoil energy ratio, as 
a function of the recoil energy. To improve the statistics, the $^{252}$Cf
data collected at -2 V and -6 V are merged. A neutron zone which
contains 95\% of the nuclear recoils is represented. The inset shows the
agreement with the CDMS results \cite{shutt} (filled squares, the
extension of which is of the order of the error bar)}
\label{fig4}
\end{figure}
\begin{figure}[ht]
\epsfig{file=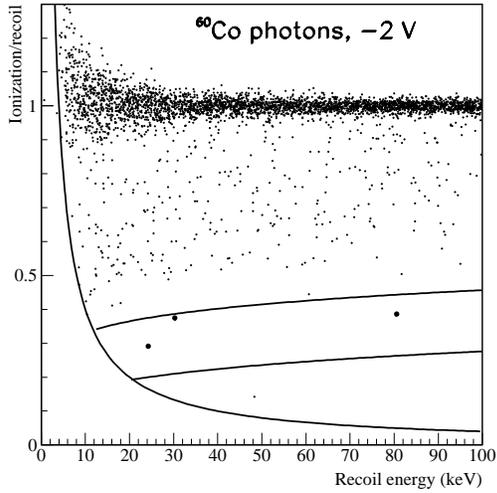,width=8cm}
\caption[] {Separation of signal from background based upon $^{60}$Co photon calibration, 
at -2 V  bias voltage. 3229 events are found over the 15 - 100 keV recoil energy interval but 
only 3 remain in the 95\% acceptance nuclear recoil band.}
\label{fig5}
\end{figure}
\begin{figure}[hb]
\epsfig{file=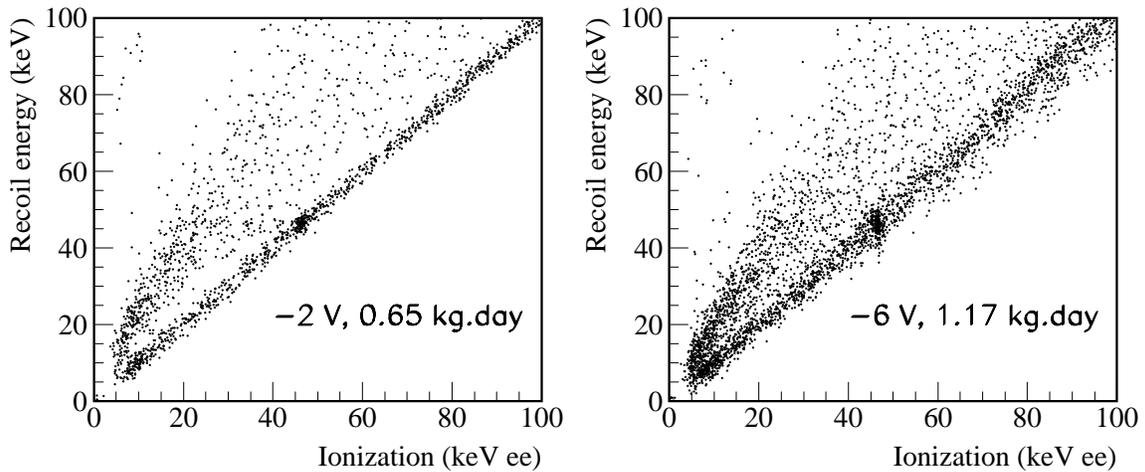,width=17cm}
\caption[] {Data at -2 V and -6 V. Deriving recoil energies from 
the heat signal by subtraction of charge times bias voltage causes the
recoil energy resolution to become bias-dependent, as can be seen on the 46.5 keV peak of $^{210}$Pb.}
\label{fig6}
\end{figure}
\begin{figure}[ht]
\epsfig{file=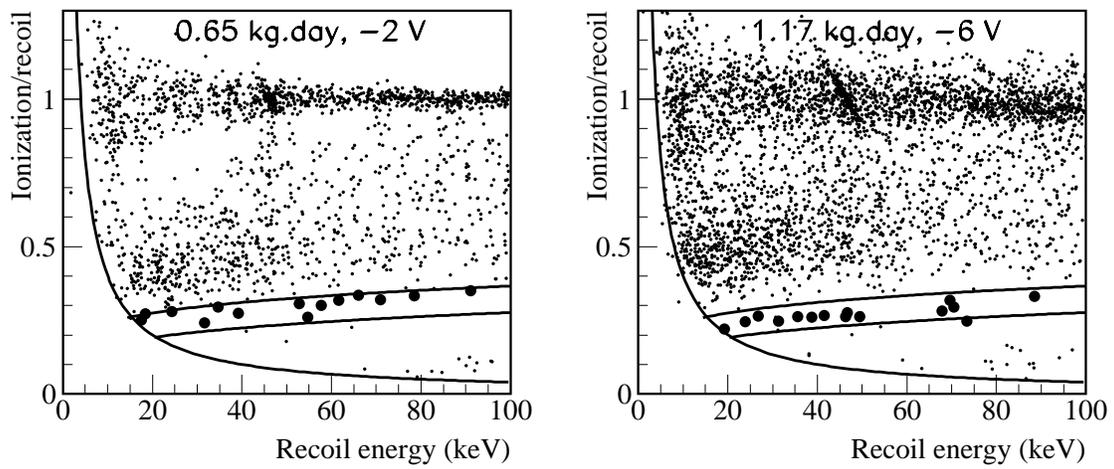,width=17cm}
\caption[] {Neutron zones and data at -2 V and -6 V. We only
retain the  (0$\sigma$, -2$\sigma$) neutron zones (47.5\% acceptance) where 14 and 15 events
(filled circles) subsist respectively over the 15-100 keV recoil energy range.}
\label{fig7}
\end{figure}

\end{document}